\documentclass[english,prd,twocolumn,superscriptaddress,nofootinbib,preprintnumbers]{revtex4}
\usepackage[latin1]{inputenc}
\usepackage{graphicx}

\makeatletter

\usepackage{float}

\makeatletter

\usepackage{dcolumn}



\makeatother

\usepackage{babel}
\makeatother
\begin{document}

\title{Power-laws $f(R)$ theories are cosmologically unacceptable}

\author{Luca Amendola}

\email{amendola@mporzio.astro.it}

\affiliation{INAF/Osservatorio Astronomico di Roma, Via Frascati 33, I-00040 Monte
Porzio Catone, RM, Italy}

\author{David Polarski}

\email{David.polarski@LPTA.univ-montp2.fr}

\affiliation{Lab.\,de Physique Math\'{e}matique et Th\'{e}orique,\\
 Universit\'{e} Montpellier II, UMR 5825, 34095 Montpellier Cedex
05, France}

\author{Shinji Tsujikawa}

\email{shinji@nat.gunma-ct.ac.jp}

\affiliation{Department of Physics, Gunma National College of Technology, Gunma
371-8530, Japan}

\date{\today{}}

\begin{abstract}
In a recent paper \cite{APT} we have shown that $f(R)=R+\mu R^{n}$
modified gravity dark energy models are not cosmologically viable
because during the matter era that precedes the accelerated stage
the cosmic expansion is given by $a\sim t^{1/2}$ rather than $a\sim t^{2/3}$,
where $a$ is a scale factor and $t$ is the cosmic time. A recent
work by Capozziello \textit{et al.} \cite{Capo} criticised our results
presenting some apparent counter-examples to our claim in $f(R)=\mu R^{n}$
models. We show here that those particular $R^{n}$ models can produce
an expansion as $a\sim t^{2/3}$ but this does not connect to a late-time
acceleration. Hence, though acceptable $f(R)$ dark energy models
may exist, the $R^{n}$ models presented in Capozziello \textit{et
al.} are not cosmologically viable, confirming our previous results
in Ref. \cite{APT}. 
\end{abstract}

\maketitle
Among the various interesting possibilities invoked in order to explain
a late-time accelerated expansion, $f(R)$ modified gravity dark energy
(DE) models ($R$ is the Ricci scalar) have attracted a lot of attention.

However, we found recently in Ref.~\cite{APT} that for a large class
of $f(R)$ DE models, including $R^{n}$ models, the usual power-law
stage $a(t)\propto t^{2/3}$ \emph{preceding} the late-time accelerated
expansion is replaced by a power-law behaviour $a(t)\propto t^{1/2}$.
Such an evolution is clearly inconsistent with observations, e.g.
the distance to last scattering as measured by CMB acoustic peaks.
Hence a viable cosmic expansion history seems to be a powerful constraint
on such models. The results in Ref.~\cite{APT} have been discussed
in greater detail and largely expanded in Ref.~\cite{APT2}. These
cosmological difficulties of $f(R)$ models are complementary to those
that arise from local gravity constraints.

However, the claim of \cite{APT} was recently criticised by Capozziello
\textit{et al.} (CNOT) \cite{Capo}. This short paper is devoted to
addressing explicitly this criticism.

It is necessary to begin with some clarifications. First, it is clear
that $f(R)$ gravity models can be perfectly viable in different contexts.
Maybe the best example is Starobinsky's model, $f(R)=R+\mu R^{2}$
\cite{star}, which has been the first internally consistent inflationary
model. This Lagrangian produces an accelerated stage \emph{preceding}
the usual radiation and matter stages. A late-time acceleration in
this model requires a positive cosmological constant (or some other
form of dark energy) in which case the late-time acceleration is not
due to the $R^{2}$ term.

Second, it is important to clarify an issue raised in CNOT concerning
the validity of the conformal transformation we used in \cite{APT}.
We checked all our results numerically (and where possible also analytically)
both in Jordan frame (JF) and Einstein frame (EF), always considering
the former as the physical frame (i.e. the frame in which matter is
conserved with an energy density $\rho_{{\rm m}}\propto a^{-3}$).
So the power-law behaviour $a\sim t^{1/2}$ found in JF is not an
artifact of the conformal transformation. It is in fact the same solution
of the original Brans-Dicke paper \cite{BD} in 1961 (Eq.~60) with
$\omega=0$ (equivalent to $\beta=1/2$ in the notation of \cite{APT})
and corresponds to solutions found in JF also in other $R^{n}$ papers
such as Ref.~\cite{barrow}.

The main criticism in CNOT is that it is possible to have a stage
with $a(t)\propto t^{2/3}$ followed by a DE dominated phase for some
$f(R)$ models and even for the power-law case $f(R)=\mu R^{n}$ (notice
that we changed the sign of $n$ with respect to our paper \cite{APT}
in order to match the choice in CNOT) and several examples are suggested.
We address here the viability of the $R^{n}$ models suggested in
CNOT.

The $f(R)$ gravity action in the JF is given by \begin{equation}
S=\int{\rm d}^{4}x\sqrt{-g}\left[\frac{1}{2\kappa^{2}}f(R)+{\mathcal{L}}_{{\rm m}}+{\mathcal{L}}_{{\rm rad}}\right]\,,\end{equation}
 where $\kappa^{2}=8\pi G$ while $G$ is the gravitational constant,
and ${\mathcal{L}}_{{\rm m}}$ and ${\mathcal{L}}_{{\rm rad}}$ are
the Lagrangian densities of dust-like matter and radiation respectively.
In the flat Friedmann Robertson Walker metric with a scale factor
$a$, we get the following equations \begin{eqnarray}
3FH^{2} & = & \kappa^{2}\,(\rho_{{\rm m}}+\rho_{{\rm rad}})+\frac{1}{2}(FR-f)-3H\dot{F}\,,\label{E1}\\
-2F\dot{H} & = & \kappa^{2}\left(\rho_{{\rm m}}+\frac{4}{3}\rho_{{\rm rad}}\right)+\ddot{F}-H\dot{F}\,,\label{E2}\end{eqnarray}
 where $F\equiv{\rm d}f/{\rm d}R$ and $H\equiv\dot{a}/a$ with a
dot being the derivative in terms of cosmic time $t$. Note that the
energy densities $\rho_{m}$ and $\rho_{{\rm rad}}$ satisfy the conservation
equations $\dot{\rho}_{{\rm m}}+3H\rho_{{\rm m}}=0$ and 
$\dot{\rho}_{{\rm rad}}+4H\rho_{{\rm rad}}=0$,
respectively.

We shall introduce the following quantities \begin{eqnarray}
x_{1}=-\frac{\dot{F}}{HF}\,,\quad x_{2}=-\frac{f}{6FH^{2}}\,,\quad x_{3}=\frac{\kappa^{2}\rho_{{\rm rad}}}{3FH^{2}}\,.\end{eqnarray}
 Then for the power-law models $f(R)=\mu R^{n}$, we obtain \begin{eqnarray}
 &  & \frac{{\rm d}x_{1}}{{\rm d}N}=-1+x_{1}^{2}-3x_{2}+nx_{2}(1+x_{1})+x_{3}\,,\label{be1}\\
 &  & \frac{{\rm d}x_{2}}{{\rm d}N}=-\frac{n}{n-1}x_{1}x_{2}+x_{2}(x_{1}+2nx_{2}+4)\,,\\
 &  & \frac{{\rm d}x_{3}}{{\rm d}N}=(x_{1}-2nx_{2})x_{3}\,,\label{be3}\end{eqnarray}
 where $N\equiv\ln a$. We also define \begin{equation}
\Omega_{{\rm m}}\equiv\frac{\kappa^{2}\rho_{{\rm m}}}{3FH^{2}}=1-x_{1}-(1-n)x_{2}-x_{3}\,.\end{equation}
 For general $f(R)$ dark energy models one needs an additional equation
to close the system \cite{APT2}. Among the fixed points which exist
for the above three dimensional system, the following three types
of solutions are important for our discussion (we assume no radiation,
i.e. $x_{3}=0$ in the following, unless otherwise stated).

\vskip 5pt \textbf{(i)} \textbf{\emph{Solution}} \textbf{A} ({}``curvature-dominated
solution\char`\"{}):

This corresponds to the fixed point \begin{equation}
(x_{1},x_{2})=\left(\frac{2(2-n)}{2n-1},\frac{5-4n}{(1-2n)(1-n)}\right)\,,\quad\Omega_{{\rm m}}=0\,.\label{pointA}\end{equation}
 The evolution of the scale factor is given by \cite{coa93} \begin{equation}
a\propto t^{\alpha_{A}}\,,\quad\alpha_{A}=\frac{(1-2n)(1-n)}{2-n}\,.\end{equation}
 It is an \emph{exact} solution \emph{in the absence of dust}, and
an asymptotic solution in the presence of dust. The latter was originally
used to give rise to a late-time acceleration for negative $n$ ({}``curvature
dominated late-time attractor'') \cite{Capo2,Carroll}. When $\alpha_{A}<0$
the expanding solution is given by $a\propto(t_{s}-t)^{\alpha_{A}}$,
which corresponds to a phantom solution.

\vskip 5pt \textbf{(ii)} \textbf{\emph{Solution}} \textbf{B} ({}``scaling
solution\char`\"{}):

This corresponds to the fixed point at which the energy fraction of
the matter does not vanish: \begin{eqnarray}
(x_{1},x_{2}) & = & \left(\frac{3(n-1)}{n},-\frac{(4n-3)}{2n^{2}}\right)\,,\nonumber \\
\Omega_{{\rm m}} & = & \frac{-8n^{2}+13n-3}{2n^{2}}\,.\end{eqnarray}
 The evolution of the scale factor is given by \begin{equation}
a\propto t^{\alpha_{B}}\,,\quad\alpha_{B}=2n/3\,.\end{equation}

\noindent \vskip 5pt \textbf{(iii)} \textbf{\emph{Solution}} \textbf{C}
({}``$\phi$ matter dominated epoch'' \cite{Luca}):

This corresponds to the fixed point \begin{equation}
(x_{1},x_{2})=\left(-1,0\right)\,,\quad\Omega_{{\rm m}}=2\,.\label{pointC}\end{equation}
 This stage is the so-called $\phi$-matter-dominated era ($\phi$MDE)
\cite{Luca} with scale factor evolution \begin{equation}
a\propto t^{\alpha_{C}}\,,\quad\alpha_{C}=1/2\,,\end{equation}
 for any $n$.

\vspace{0.3cm}
 It was shown in \cite{APT} that for all $n$ the $\phi$MDE replaces
the usual matter era prior to the late-time acceleration driven by
the point A. CNOT instead pointed out that it is possible to use either
solution A or B in order to have a standard matter era ($a\propto t^{2/3}$)
followed by an accelerated expansion. Clearly, two possibilities arise:
either the universe goes from A to B or vice versa. In the first case
the solution A has to behave as a matter era ($\alpha_{A}=2/3$),
and therefore $n=-0.129$ or $n=1.295$. In the second case we require
the condition $\alpha_{B}=2/3$, which corresponds to $n=1$. Hence
the three possible {}``counter examples'' suggested by CNOT are:
$n=-0.129$, $n=1.295$ and $n=1$. Now we shall investigate whether
these cases really provide a viable cosmological evolution.

Let us first analyse the stability of the solutions A and B. Neglecting
radiation and considering linear perturbations around the fixed points
$(x_{1},x_{2})$ along the line presented in Ref.~\cite{CST}, we
obtain the following eigenvalues of the corresponding Jacobian matrix
evaluated at the each fixed point: \begin{equation}
\mu_{1}=-\frac{5-4n}{1-n}\,,\quad\mu_{2}=-\frac{8n^{2}-13n+3}{(1-2n)(1-n)}\,,\end{equation}
 for the solution A, and \begin{equation}
\mu_{\pm}=\frac{3(1-n)\pm\sqrt{(1-n)(-256n^{3}+608n^{2}-417n+81)}}{4n(n-1)}\,,\end{equation}
 for the solution B.

This shows immediately that the case $n=-0.129$ (and values in the
vicinity) is excluded because the point A is then stable ($\mu_{1}<0$,
$\mu_{2}<0$) and, once reached, it will never give way to a late-time
acceleration. In other words, the transition from A to B is impossible
in this case. When $n=1.295$, A is a saddle point ($\mu_{1}<0$,
$\mu_{2}>0$) and B is a stable spiral. Hence the the transition from
A to B is possible, but with this value of $n$ (and values in the
vicinity) the point B is not accelerated, since then $\alpha_{B}=0.863$.
It is also important to note that the point A corresponds to a solution
without matter ($\Omega_{{\rm m}}=0$), so this would be a {}``matter
era'' without matter, which is clearly not acceptable as well. This
leaves as the only possibility $n=1$ and a transition from B to A.

{}From Eq.~(\ref{pointA}) the point A disappears for $n=1$, which
means that the transition from B to A is not possible. As this case
merely corresponds to Einstein gravity, it is obvious that one gets
the required behaviour $a\propto t^{2/3}$ in the dust-dominated era.
However there is no mechanism left for the generation of a late-time
acceleration unless some additional DE component is introduced, which
is what modified gravity DE models are supposed to avoid.

So we conclude from the discussion above that the solutions suggested
in CNOT do not lead to a $a\propto t^{2/3}$ behaviour followed by
an accelerated expansion.

Still it may be interesting to consider the scenario with $n$ close
to 1, for which $a\propto t^{p}$ with $p\approx2/3$, instead of
exactly $2/3$. Let us study the case with $n$ in the conservative
range $0.75<n<1.25$, which corresponds to power-law exponents $1/2<p<5/6$.
Since the point B is a stable spiral for $1<n<1.327$, transition
from a decelerated matter era to an accelerated one is impossible
also in this case. For $0.713<n<1$ the point B is a saddle, so a
transition is indeed possible. For these values the point A corresponds
to a stable node with an effective phantom equation of state \begin{equation}
w_{{\rm eff}}=-1+\frac{2(2-n)}{3(1-2n)(1-n)}<-7.6\,.\label{weff}\end{equation}
 Meanwhile the third point C, the $\phi$MDE, is a stable point as
well with an effective equation of state $w_{{\rm eff}}=1/3$. Then
the trajectories leaving the point B are attracted either by A or
C. However the final accelerated point A is a strongly phantom one
as given in Eq.~(\ref{weff}). In addition the more one tries to
obtain the exact matter era ($n\to1$) the more phantom the final
acceleration is ($w_{{\rm eff}}\to-\infty$ as $n\to-1$). Hence from
this point of view these models are cosmologically unacceptable.

\begin{figure}
\includegraphics[scale=1.0]{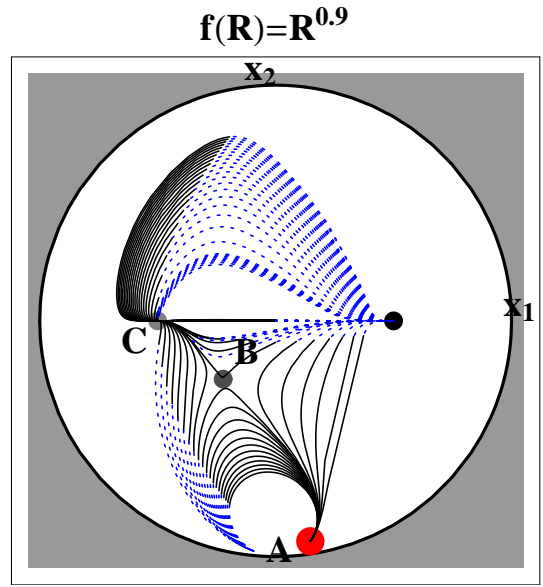} 
\caption{\label{ps09} Phase space plot in the plane $(x_{1},x_{2})$ in Poincaré
coordinates for the model $f(R)=\mu R^{0.9}$ in the \textit{absence}
of radiation. The dotted lines represent trajectories at the early
stage, whereas the continuous lines correspond to those at the final
stage. We show the fixed points A, B and C as circles. The final attractor
is either the curvature-dominated solution A or the $\phi$MDE point
C. In the absence of radiation it is possible to have the sequence
from the (quasi) matter point B to the phantom point A with restricted
initial conditions. As $n$ gets closer to 1, however, the matter
era becomes very short and the effective equation of state for the
point A diverges ($w_{{\rm eff}}\to-\infty$). If we take into account
radiation (corresponding to initial conditions $|x_{1}|\ll1$, $|x_{2}|\ll1$
and $x_{3}\approx1$) the solutions do not approach the (quasi) matter
point B before reaching the final attractor, which means the matter
epoch is absent. }
\end{figure}

To complete our proof, we have run our numerical code to investigate
the evolution of the system in the space $(x_{1},x_{2})$. Without
including radiation the final attractor is in fact either A or C depending
upon initial conditions. However trajectories which are attracted
to B first and then finally approach the point A are restricted in
a narrow region of phase space (see Fig.~\ref{ps09}). Moreover the
duration of the (quasi) matter epoch gets shorter as we choose the
values of $n$ closer to 1 (which is in fact required to obtain the
matter phase). This is associated with the fact that the eigenvalues
of the matrix of perturbations diverge in the limit $n\to1$ \cite{APT2}.
When we start from realistic initial conditions ($|x_{1}|\ll1,|x_{2}|\ll1$)
with inclusion of radiation, the solutions directly approach the fixed
point A or C without passing the vicinity of the point B. In other
words we have not found any trajectories in which the radiation era
is followed by matter and final accelerated eras. Therefore, our numerical
analysis excludes also the range $0.713<n<1$ as a viable cosmological
model.

In CNOT the possibility is also mentioned of reconstructing the theory
from observations (in particular from the function $H(z)$ where $z$
is the redshift), in analogy to the reconstruction of scalar-tensor
DE models \cite{BEPS00,EP01}. Clearly we expect that many $f(R)$ DE models
can be successfully reconstructed \emph{at low redshifts} for any
$H(z)$ corresponding to late-time accelerated expansion. However,
nothing guarantees that an $H(z)$ \emph{corresponding to a conventional
cosmic expansion at high $z$} leads to an acceptable form of $f(R)$. 
Sometimes a singular behaviour arises already at 
very low redshifts \cite{EP01,AT}.
The procedure of reconstruction does not guarantee either the stability 
of the solution on higher redshifts. Finally, the particular reconstructed 
model proposed in CNOT does not contradict our claim on $f(R)=\mu R^{n}$ 
or $f(R)=R+\mu R^{n}$ ($n\neq0$) models since the reconstruction attempted 
in CNOT is not for $f(R)$ models of this type.

The question of the cosmological viability of $f(R)$ models is a
very interesting one. In Ref. \cite{APT2} we have spelled out the
conditions on the forms of $f(R)$ in order to satisfy the basic cosmological
requirements of a standard matter era and a late-time accelerated
attractor. The specific $f(R)=\mu R^{n}$ examples suggested in CNOT
do not pass these criteria and therefore confirm, rather than contradict,
our claim.

%%%%%%%%%%%%%%%%%%%%%%%%%%%%%%%%%
\section*{ACKNOWLEDGMENTS}
We are grateful to R.~Gannouji and A.~A.~Starobinsky for useful
discussions. We also thank B.~Bassett, S.~Capozziello, S.~Nojiri,
S.~Odintsov and R.~Tavakol for useful correspondence. S.~T. is
supported by JSPS (Grant No.\,30318802). 

%%%%%%%%%%%%%%%%%%%%%%%%%%%%%%%%%

\end{document}